# Fisheye lens distortion correction

Dmitry Pozdnyakov

*Oxagile, Dzerginskogo av.57, Minsk 220089, Belarus*

*E-mail: pozdnyakov@tut.by; dmitry.pozdnyakov@oxagile.com*

**Abstract:** A new distortion correction algorithm for fisheye lens with equidistant mapping function is considered in the present study. The algorithm is much more data lossless and accurate than such a classical approach like Brown–Conrady model.

### Introduction

At present cameras with fisheye lens are spread very much. It is because the cameras like that has the field of view up to or even more than half a complete solid angle (horizontal and vertical viewing angle is equal or more than 180°) that provides obvious advantages in comparison with solution based on several cameras or camera on a pivot bracket. At the same time the images (frames) from such cameras have significant geometrical distortion. So quite often it is necessary to apply digital post-processing of the images to compensate their distortion for information perception convenience. In particular, there is Brown–Conrady method [1, 2] that has already become a classical approach. Meanwhile new and new approaches for the problem solution are developed. And this is due to the following circumstances. The fact is that the complete correction of the distortion is associated with corresponding image projection from semi-sphere or sphere-like surface to the plane. But in this case under the condition that the spatial resolution is preserved in the central area of the image the size of the output straightened image increases dramatically. Because of that the necessity to cut the output image around the perimeter arises and the main advantage of the fisheye lens (field of view) is partly lost. Since it is fundamentally impossible to satisfy simultaneously three contradicting requirements (non-reduction of the spatial resolution of the image, non-increase of the image size in pixels and complete compensation of the geometric distortion), the development of more and more balanced models is a very important issue (see, for example, [3 –6]) and it will remain so in the future.

Thus the purpose of the manuscript is development of a new effective algorithm of geometrical distortion correction for fisheye lens with equidistant mapping function and field of view equal or less than half a complete solid angle.

### Theory and results

Let us consider one of such mostly used fisheye lens optical systems as equidistant scheme [7] that has only radial component of the barrel distortion [8] in case of application of the high quality optical elements. Its correction in a more general form than classical representation through series by $r$ [1, 2, 8] can be done by means of equality

$$\mathbf{r}_s = \mathbf{r}_p f(r_p). \quad (1)$$

This equality associates the brightness value (or three brightness values) of pixel characterized by the radius vector $\mathbf{r}_p$ measured from the center of corrected monochrome (or color) image with the brightness value (values) of pixel characterized by the radius vector $\mathbf{r}_s$ measured from the center of the original initial image. That is, Eq.(1) set a rule for recalculation of pixels (with coordinates $(x_p, y_p)^T = \mathbf{r}_p$) belong to a plane through corresponding pixels (with coordinates $(x_s, y_s)^T = \mathbf{r}_s$) belong to a semi-sphere by means of application of the mapping function $f$.

The rigorous analytic expression can be obtained for the mapping function $f$ in the considered case. It has the following view:

$$f(r) = \frac{2R_0}{\pi r} \operatorname{arctg}\left(\frac{\pi r}{2R_0}\right). \quad (2)$$

Here $R_0 = 2r_0$, $r_0$ ($R_0$) is the radius (in pixels) corresponding to angle $\pi/4$ ($\pi/2$) rad on the initial image relative to the lens axis symmetry. This parameter is unique for each concrete optical system. Obviously Eq.(2) can be expanded into Taylor series with both consideration only several series terms and renormalization of their coefficients to represent the function $f$ in standard form according to Brown–Conrady model [1, 2, 8]. But further just the rigorous Eq.(2) for function $f$ will be considered.

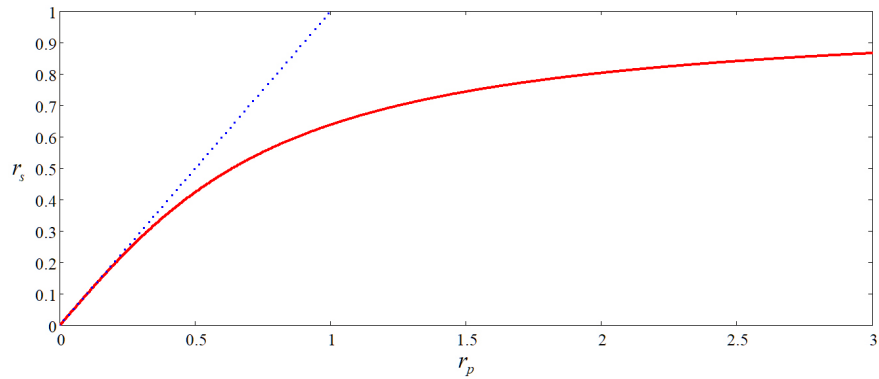

**Fig.1** – Coordinates transformation function $r_p \to r_s$ according to Eq.(2) at $R_0 = 1$

In accordance with Fig.1 under the condition that both the spatial resolution (hereinafter this condition is for a central part of the image) and image sizes are preserved the angle of view is reduced from 180° to 115° due to distortion correction (see result on Fig.2).

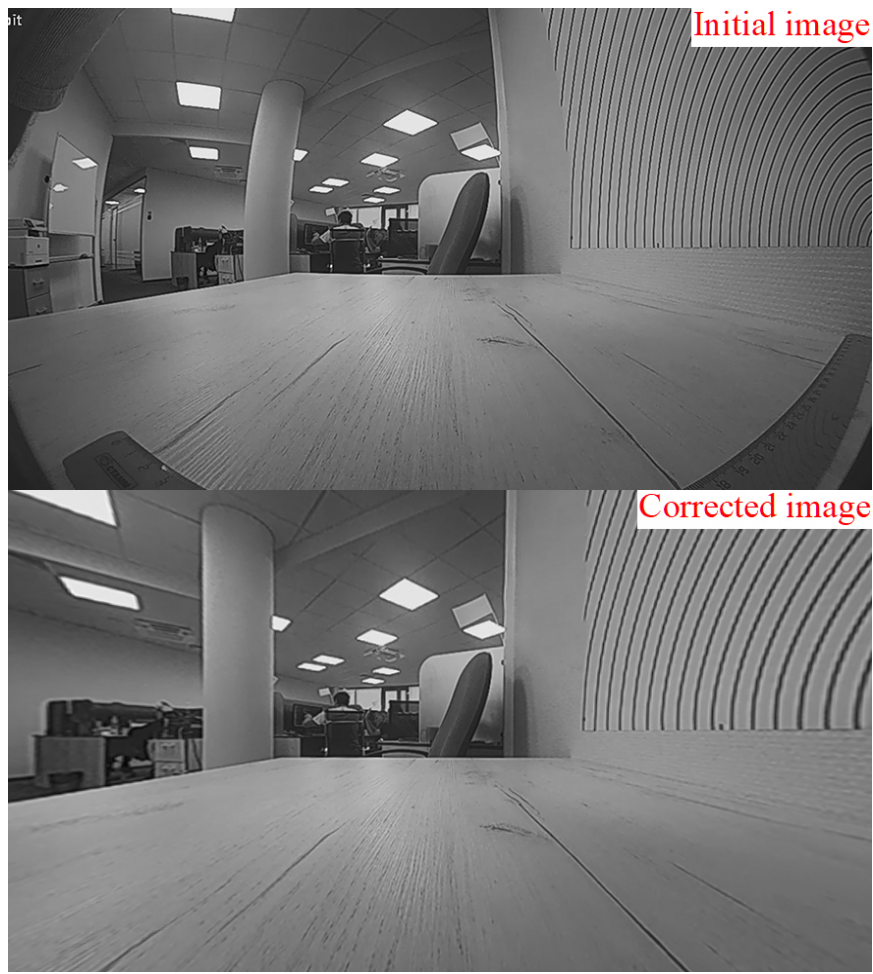

**Fig.2** – Complete distortion correction according to Eq.(2) at $(r_p/R_0) \in [0,1]$

If to increase the area of output image in four times with preserving the spatial resolution then the angle of view increases up to 145° (see result on Fig.3).

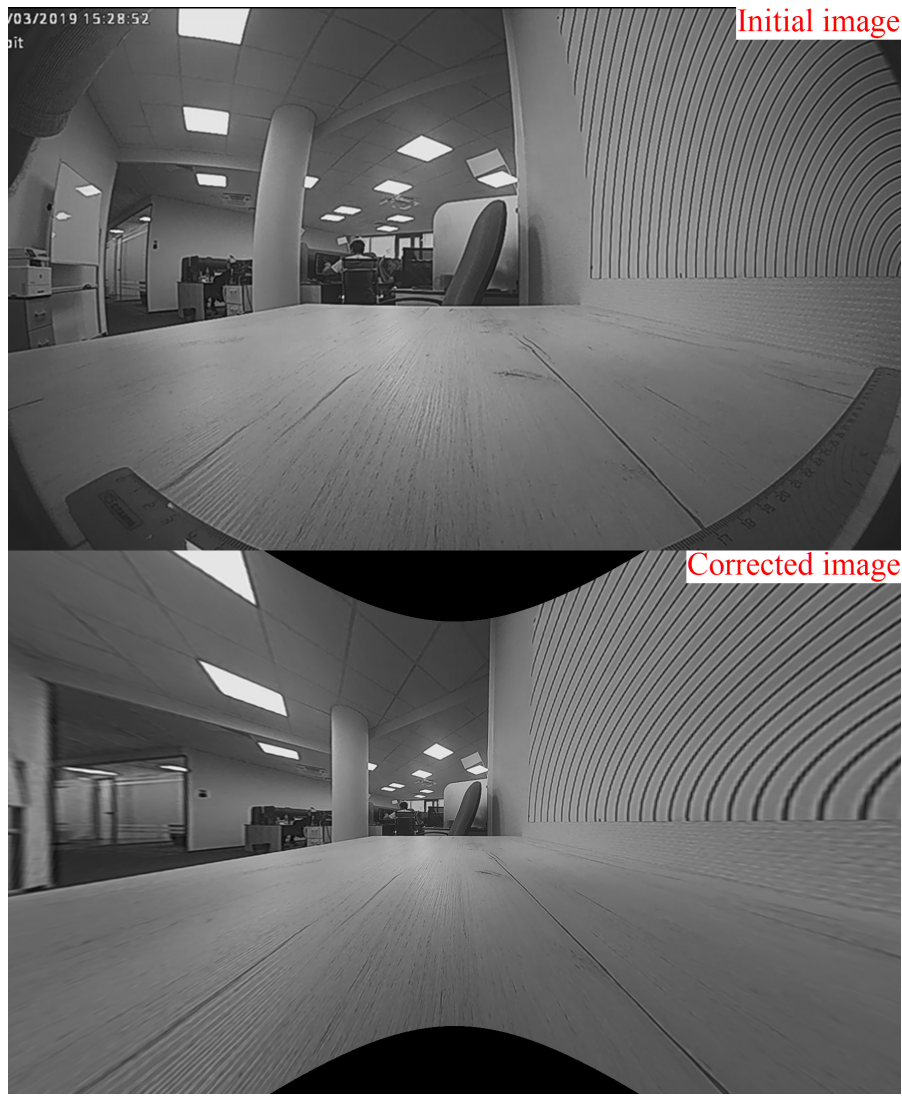

**Fig.3** – Complete distortion correction according to Eq.(2) at $(r_p/R_0) \in [0, 2]$

Further giant increase of the output image sizes does not provide significant increase of the angle of view. To solve the contradiction it is necessary to modify the mapping function $f \to F$ in such a way that, from one hand, it is still very close to the initial one in the vicinity of zero and, from other hand, it is maximally linear in the vicinity of unity to exclude the second order distortions. The conditions like that can be satisfied properly for the following mapping-function:

$$F(r) = \frac{2R_0}{\pi r} \arctg\left(\frac{4\pi R_0^2 r}{8R_0^3 - r^3}\right) + \frac{2R_0}{r} \Theta\left(\frac{r}{2R_0} - 1\right), \quad (3)$$

$$\begin{cases} \Theta(x) = 0, \ x \leq 0, \\ \Theta(x) = 1, \ x > 0. \end{cases} \quad (4)$$

Although application of Eq.(3) instead of Eq.(2) allows the wide angle of view (180°) to be preserved under the conditions that the spatial resolution is the same and the image area is larger in four times or the image sizes are the same and the spatial resolution is lower in two times, but such a modification of the mapping-function does not allow the barrel distortion to be eliminated in the peripheral part of the image (see Fig.4).

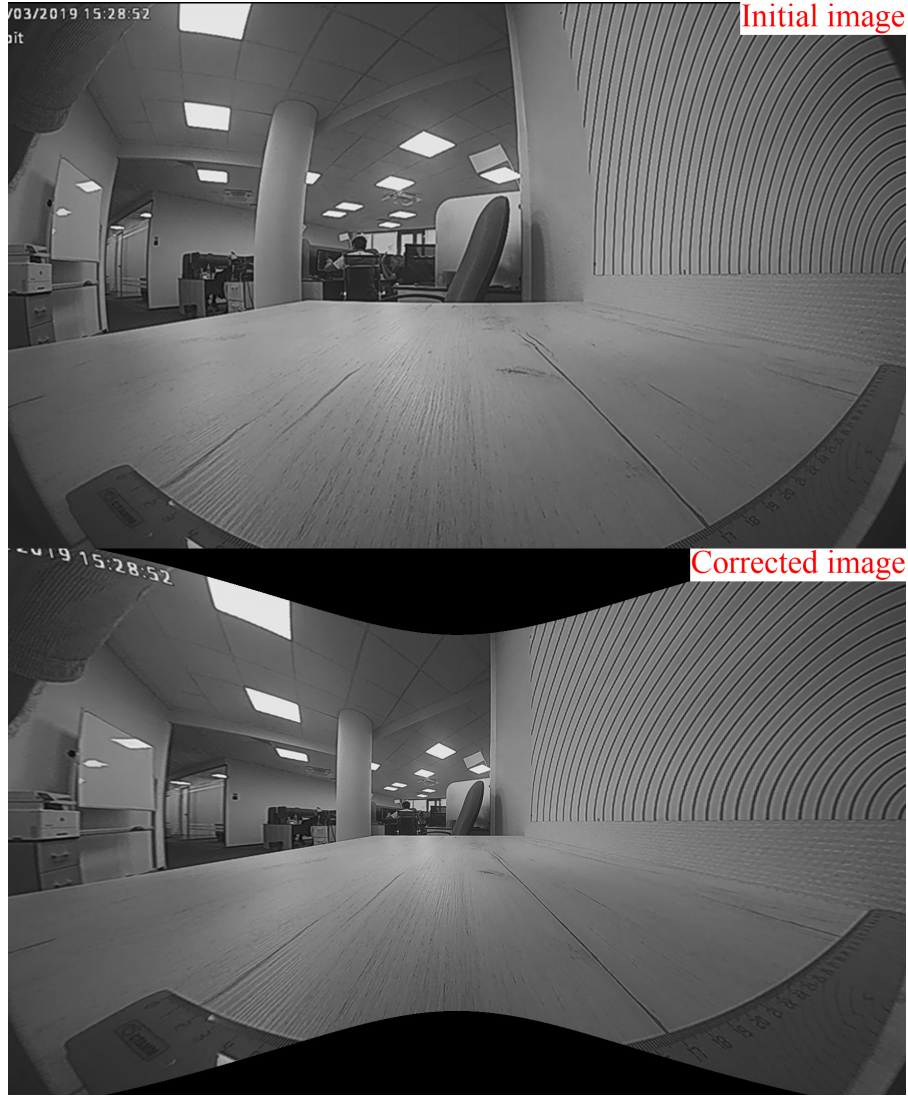

**Fig.4** – Incomplete correction of the distortion according to Eq.(3) at $(r_p/R_0) \in [0,2]$

To correct the barrel distortion of the image peripheral part the significantly non-linear deformations of circles into squares with smoothed corners depending on **r** should be used (in particular, the circle inscribed into the outer perimeter of the image should deform into the square directly). There are expressions could be used for corresponding pixel-to-pixel transformation below

$$\mathbf{r}_t = \mathbf{r}_p G\left(\frac{x_p^2 y_p^2}{r_p^2}, S\left(\frac{x_p^2}{4R_0^2}, \frac{y_p^2}{4R_0^2}, \frac{r_p^2}{8R_0^2}\right)\right), \quad (5)$$

$$G(w,z) = \left\{ \frac{4w\cos(\pi z/2)}{1+\cos(\pi z/2) - \left[1+\cos(\pi z/2)+(1-8w)\left(\cos(\pi z/2)+\cos^2(\pi z/2)\right)\right]^{1/2}} \right\}^{1/2}, \quad (6)$$

$$S(h,v,p) = 1 - \left(h^{1+\text{tg}(\pi p/2)} + v^{1+\text{tg}(\pi p/2)}\right)^{\frac{3/2}{1+\text{tg}(\pi p/2)}}. \quad (7)$$

The result of application of Eq.(5) is presented in Fig.5. As follows from the figure the arcs transform to the horizontal and vertical straight lines near the image edges. Distortion of the image central region still remains but its value partly decreases.

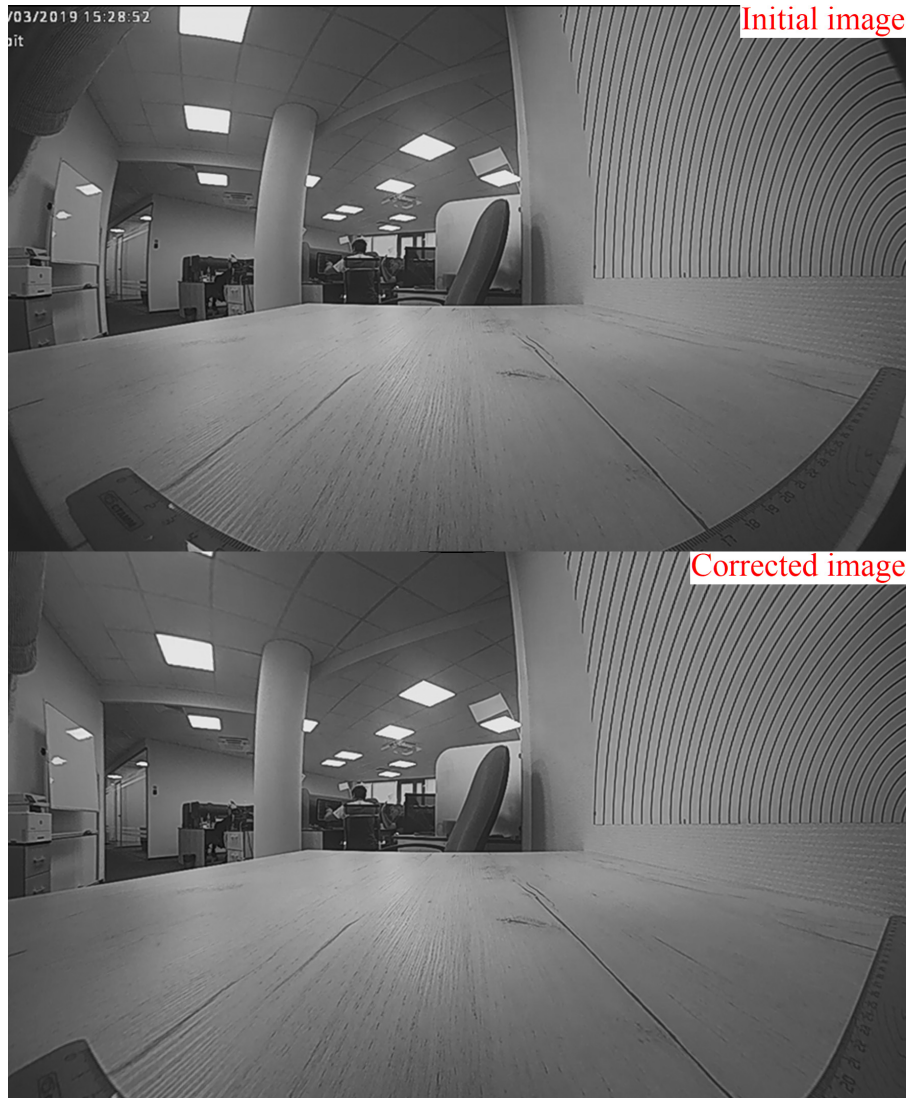

**Fig.5** – Preliminary correction of the image according to Eq.(5) at $(r_p/R_0) \in [0,2]$

For the subsequent correction of the residual distortion it is also necessary to apply Eqs.(1) and (3) to the intermediate result, obtained by means of Eq.(5), in the standard form

$$\mathbf{r}_s = \mathbf{r}_t F(r_t). \quad (8)$$

The figures below show the final results of successive image transformations according to Eqs.(5) and (8). As it follows from the figures the barrel distortion is very effectively corrected over all of the field of view with preserving both the wide angle of view (180°) and the spatial resolution with increase of the image area only in four times.

In Fig.7 the camera axis was specially aligned with the center of the printed circles of different radius and thicknesses so that both the distance between the circles and their thickness were visually the same for the fisheye lens with equidistant optical system.

In all represented figures the black regions correspond to the transformation $\mathbf{r}_s \leftrightarrow \mathbf{r}_p$ when the calculated radius-vector $\mathbf{r}_s$ is out of the initial image.

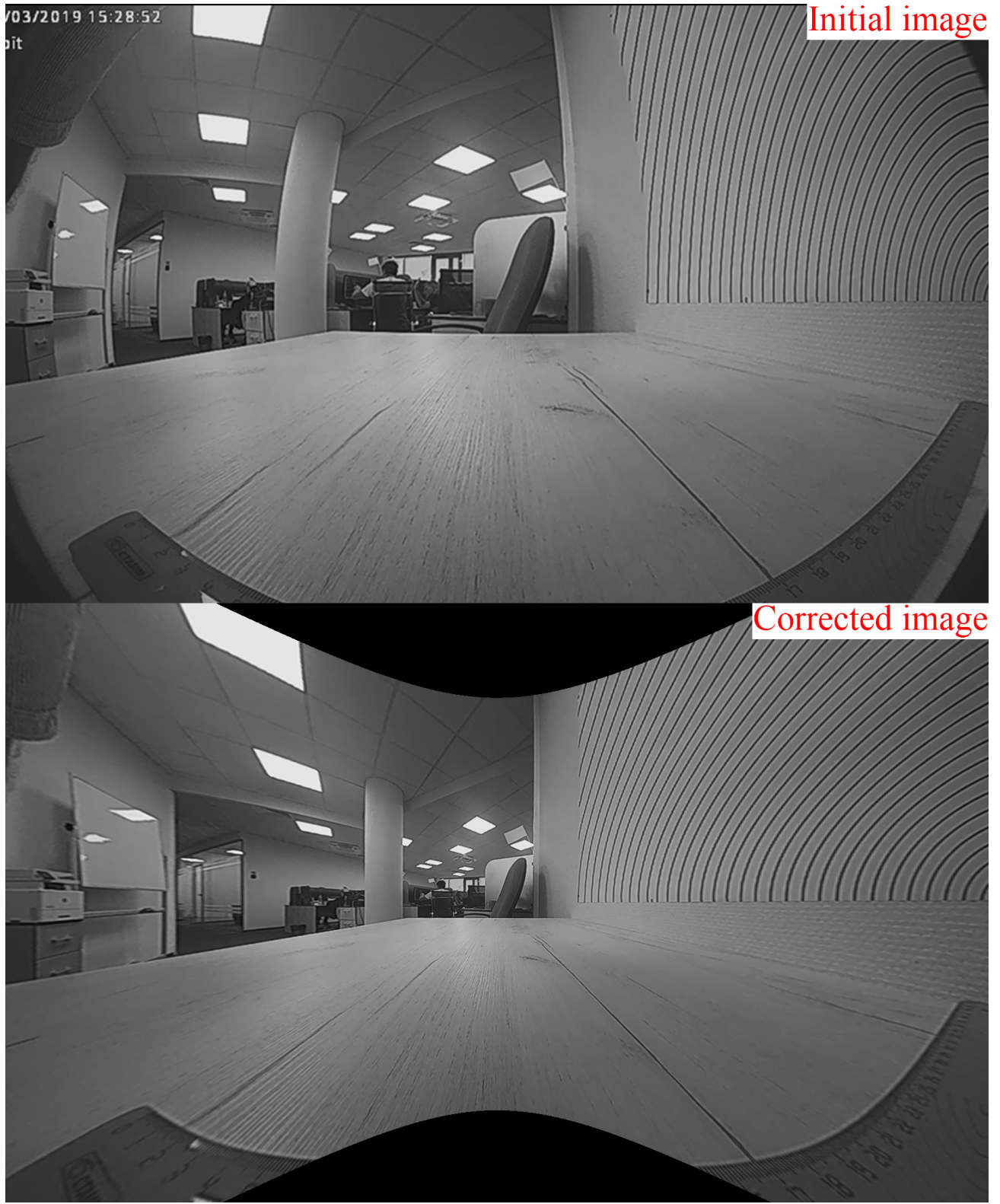

**Fig.6** – Distortion correction of the first image

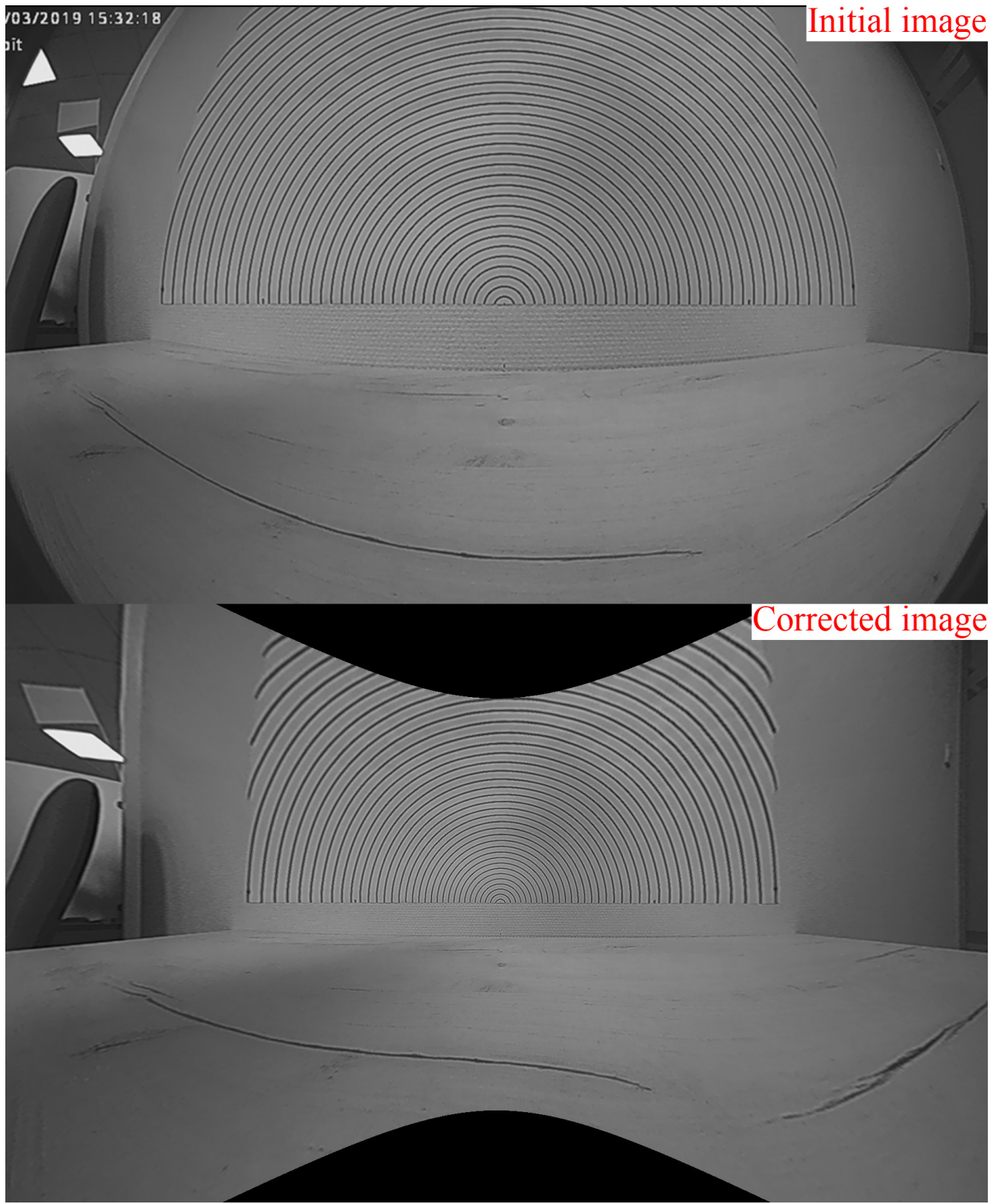

**Fig.7** – Distortion correction of the second image

**Conclusion**

The new effective algorithm of distortion correction for fisheye lens with equidistant mapping function has been developed. It allows the spatial resolution along with a wide viewing angle to be preserved with the minimal increase of the image or frame sizes.